\documentclass[lettersize,journal]{IEEEtran}
\usepackage{amsmath,amsfonts}
\usepackage{array}
\usepackage[caption=false,font=normalsize,labelfont=sf,textfont=sf]{subfig}
\usepackage{textcomp}
\usepackage{stfloats}
\usepackage{url}
\usepackage{verbatim}
\usepackage{graphicx}
\usepackage{cite}
\usepackage{xcolor}
\usepackage{footmisc}
\usepackage{multirow}
\usepackage{float}
\usepackage{booktabs} 
\usepackage[linesnumbered,ruled,noend]{algorithm2e}
\usepackage{caption}
\usepackage{hyperref}
\begin{document}
\setlength{\floatsep}{5pt}       
\setlength{\textfloatsep}{10pt}   
\setlength{\intextsep}{5pt} 
\title{CONCERTO: Complex Query Execution Mechanism-Aware Learned Cost Estimation}

\author{
\IEEEauthorblockN{
Kaixin Zhang,
Hongzhi Wang\textsuperscript{*}\thanks{Corresponding author},
Kunkai Gu,
Ziqi Li,
Chunyu Zhao,
Yingze Li,
Yu Yan,
}
\IEEEauthorblockA{
\\Harbin Institute of Technology\\
Email: 21B903037@stu.hit.edu.cn, wangzh@hit.edu.cn, {24S003033, 24B903030, 24B903032, 23B903046}@stu.hit.edu.cn, yuyan@hit.edu.cn
}
}


\markboth{Journal of \LaTeX\ Class Files,~Vol.~14, No.~8, August~2021}%
{Shell \MakeLowercase{\textit{et al.}}: A Sample Article Using IEEEtran.cls for IEEE Journals}


\maketitle

\begin{abstract}
With the growing demand for massive data analysis, many DBMSs have adopted complex underlying query execution mechanisms, including vectorized operators, parallel execution, and dynamic pipeline modifications. However, there remains a lack of targeted Query Performance Prediction (QPP) methods for these complex execution mechanisms and their interactions, as most existing approaches focus on traditional tree-shaped query plans and static serial executors. To address this challenge, this paper proposes CONCERTO, a Complex query executiON meChanism-awaE leaRned cosT estimatiOn method. CONCERTO first establishes independent resource cost models for each physical operator. It then constructs a Directed Acyclic Graph (DAG) consisting of a dataflow tree backbone and resource competition relationships among concurrent operators. After calibrating the cost impact of parallel operator execution using Graph Attention Networks (GATs) with additional attention mechanisms, CONCERTO extracts and aggregates cost vector trees through Temporal Convolutional Networks (TCNs), ultimately achieving effective query performance prediction. Experimental results demonstrate that CONCERTO achieves higher prediction accuracy than existing methods.
\end{abstract}


\section{Introduction}
\label{section:Introduction}

\IEEEPARstart{O}{nline} Analytical Processing (OLAP) databases are a type of database system designed specifically to support complex queries and large-scale data analysis. With the development of OLAP DBMSs, there were already many advanced execution mechanisms proposed to improve OLAP DBMSs' execution speed, such as applying the bulk (chunk) processing to speed up queries used by PostgreSQL and MonetDB~\cite{MonetDB}. Some advanced OLAP DBMSs such as ClickHouse~\cite{clickhouse}, SparkSQL~\cite{SparkSQLAQE}, Databricks~\cite{AQE}, and DuckDB~\cite{DuckDB} pushed the OLAP query processing technology to a new level by introducing a dynamic pipeline instead of using a conventional static query execution engine. Some of them, such as ClickHouse and SparkSQL, go a step further and use a highly parallel DAG-based pipeline structure instead of the tree-shaped physical plan executed by a Volcano-style engine. Such huge and unique modifications to the query plan model and engine lead to many new demands for components like the cost estimator and query optimizer, etc. As shown in \autoref{tab:summary}, we refer to OLAP DBMSs that support all these complex execution mechanisms as \textit{high-performance OLAP DBMSs} and focus on the learned QPP method toward them.

\begin{table}[htbp]
\centering
\caption{Comparison of mechanisms supported by DBMSs. S: support, N: not-support, PS: partial-support, IS: indirect-support, AQE: adaptive query execution, and Dynamic: dynamically changing operators and pipelines. Exe. is the abbreviation for Execution.}
\label{tab:summary}
\begin{tabular}{lllll}
\toprule
           & SIMD                      & Batch Exe.                        & Parallel Exe.                  & Dynamic Exe. \\ \midrule
PostgreSQL & PS                        & PS                                & PS                             & N        \\         
MonetDB    & PS                        & S                                 & S                              & N       \\
SparkSQL   & IS                        & S                                 & S                              & AQE      \\
Databricks & IS                        & S                                 & S                              & AQE      \\
DuckDB     & S                         & S                                 & S                              & Dynamic  \\
ClickHouse & S                         & S                                 & S                              & Dynamic  \\
\bottomrule
\end{tabular}
\end{table}

Query performance prediction is a widely needed task in admission control~\cite{QCop, AreOCMUnusable}, resource management~\cite{AreOCMUnusable, SLATree, STeP}, and query optimization~\cite{NEO, ASurvey, SparkSQLAQE, AQE}, which mainly aims to estimate the latency of queries. The accuracy of QPP directly impacts all the above data management tasks and is crucial to the performance and stability of the DBMS. QPP can be regarded as a specific type of cost estimation problem that uses query latency to measure query cost. Before the widespread introduction of deep learning into the database domain, mainstream databases primarily employed conventional methods to predict query plan performance. These methods mostly relied on simple linear models, combining the cost coefficients of physical operators such as sequential scan and indexing with the input scale provided by cardinality estimators to derive query performance. Although such conventional methods are widely used in some advanced OLAP DBMSs such as SparkSQL~\cite{SparkSQLAQE} and Databricks~\cite{AQE}, they ignore the inherent complexity of DBMS and can only achieve limited accuracy. For example, the single operator's cost can be variable under different query pipeline contexts and different system workloads due to the cache changing and resource utilization caused by other operators, especially under the dynamic pipeline environment. Such complexity is even more significant in some OLAP DBMSs, which take the high degree of parallelism of query pipelines as the core design idea. Solving such complex problems is exactly what the learning cost estimator is good at.

With the development of deep learning, numerous studies have applied deep learning models to the QPP field, aiming to directly measure the quality of query plans based on the time cost of query execution. Existing works primarily focus on feature representation and end-to-end prediction for tree-shaped query plans. For instance, in databases like PostgreSQL~\cite{PostgreSQL} where query plan trees serve as physical plans, many methods~\cite{NEO, End2End, QPPNet} have been demonstrated to achieve higher accuracy and prediction speed. In the field of concurrent query performance prediction, there have also been studies~\cite {GraphEmbQPP} treating concurrent queries' plan trees as Directed Acyclic Graphs (DAGs) and treating resource competition and data sharing relationships between concurrent queries as edges that connect those sub-DAGs to capture resource competition relationships among a group of operators. 

However, most existing learning-based methods are designed for DBMSs with tree-shaped query plans and static serial pipelines. They lack sufficient targeted design and awareness of the underlying query execution mechanisms in DBMSs, particularly for mechanisms commonly used in OLAP databases like vectorized operators, Directed Acyclic Graph (DAG) pipelines, parallel query execution engines, dynamic pipeline modifications (or adaptive query execution), as well as the synergistic effects of these mechanisms.

\textbf{\textit{C.1. Difficulty in capturing the synergistic effects of complex execution mechanisms.}} An increasing number of database systems~\cite{clickhouse, impala,MonetDB} employ SIMD instruction sets (such as SSE and AVX~\footnote{\url{https://clickhouse.com/blog/cpu-dispatch-in-clickhouse}}) to implement vectorized physical operators. These operators, due to their higher computational throughput, exert greater pressure on resources like CPU caches and memory I/O. When combined with mechanisms like parallel query execution, this makes it even more challenging for models to capture the key characteristics of query execution in an end-to-end manner, thereby limiting the accuracy of cost estimation.

\textbf{\textit{C.2 Dynamic pipeline extraction and expression.}} Some DBMSs~\cite{SparkSQLAQE, AQE, clickhouse} employ runtime optimization to alleviate issues such as inaccurate estimation of key information, such as cardinality, before querying, or changes in data distribution and system load during runtime. Such a mechanism will add extra operators during execution and change the operator algorithms, such as sort and join. In summary, there are significant differences between such high-performance OLAP DBMSs and other databases, and existing QPP methods are not designed for them. Such a mechanism makes it inaccurate to use the pipeline extracted before the query, and it is also difficult to express any dynamic pipeline extracted in a form that can be easily processed by learned QPP methods. It's difficult to extract a low-level runtime dynamic pipeline that is detailed enough to support the QPP task. Also, even if such information can be extracted, it is still difficult to process with low cost by learned QPP methods, especially when the pipeline is in spatiotemporal graph form since the inference on spatiotemporal graph networks is expensive.

\textbf{\textit{C.3 Parallel execution and resource competition. }} Some DBMSs~\cite{SparkSQLAQE, clickhouse} employ parallel pipeline executors, which introduce more complex resource competition within a single OLAP query. Taking ClickHouse~\footnote{\url{https://github.com/ClickHouse/ClickHouse/issues/34045}} as a representative example, its parallel execution of ClickHouse consists of two parts, horizontal and vertical parallel. The horizontal parallel execution means that different parts of the data are processed in parallel by the same computation. ClickHouse achieves this by widely using SIMD instruction sets like SSE and AVX~\footnote{\url{https://clickhouse.com/blog/cpu-dispatch-in-clickhouse}} and creating multiple execution branches (threads) for operators like \textit{MergeSorting} to speed up queries. Such operators' costs are highly related to the system workload and much harder to predict within an end-to-end model. The vertical parallel execution means that different stages of computation are run in parallel. To fit such a design and speed up queries, different from the widely used volcano model, these DBMSs' query plans are constructed as parallel pipelines\footnote{\label{footnote:pipeandDAG}\url{https://presentations.clickhouse.com/meetup24/5.\%20Clickhouse\%20query\%20execution\%20pipeline\%20changes/}}, resulting in significant resource competition among operators within a single query plan, including competition for CPU, I/O, and memory resources. Existing methods focus on common databases with tree-like query plans, how to capture the resource competition information, and calibrate operators' costs still remain to be studied.

In this paper, we propose a multi-stage QPP method, CONCERTO, to tackle the challenges above and give an accurate and efficient prediction, which decouples the operator cost prediction, cost calibration, and query performance prediction. The core idea of CONCERTO to achieve high accuracy for high-performance OLAP DBMSs is first to obtain detailed low-level physical pipeline information that includes dynamic pipeline changes, and then to use multi-stage models to predict the resource cost of SIMD operators separately, calibrate the resource competition, and summarize the overall pipeline latency. 

For challenge \textit{\textbf{C.1}}, CONCERTO decouples operator cost prediction from query performance prediction. We establish lightweight cost models (a.k.a \textsf{operator cost predictors (OCPs)}) for each operator to predict the resource and time cost of each operator's execution as \textbf{Stage 1}. These individual operator cost vectors are then aggregated into a graph for subsequent processing.

For challenge \textit{\textbf{C.2}}, CONCERTO includes a \textsf{Runtime Tracker} to extract training data and necessary dynamic information for prediction. To collect training data of the \textsf{OCPs}, CONCERTO uses a novel serial executor, and the tracker runs in full-collection mode that marks all data chunks by their addresses. For query-level training data of training and prediction, considering high-performance OLAP DBMSs' dynamic parallel execution mechanism and inspired by ClickHouse's probe phase~\cite{clickhouse}, the tracker uses a probe execution mode which only launches a few chunks to obtain operator features and the execution paths, which usually contain more operators than the initial pipeline. For dynamic pipeline expression, CONCERTO uses the data-flow tree rather than the pipeline DAG as the backbone of the data structure. CONCERTO includes a \textsf{Graph Constructor} to build the resource competition graph with the backbone data-flow tree and corresponding operators' cost vectors based on the execution paths of chunks collected by the \textsf{Runtime Tracker}. Therefore, the chunks' execution information generated by the modified pipeline can be reserved in different execution paths of the data-flow tree without mixing up.

For challenge \textit{\textbf{C.3}}, we introduced resource competition into the pipeline and designed an attention-based resource competition intensity adaptive mechanism and then used Graph Attention Network(GAT)~\cite{GAT} to calibrate the cost as \textbf{Stage 2}. In particular, CONCERTO categorizes operators based on their resource usage and assigns them to different resource competition groups. We establish edges in the generated data-flow tree based on those resource competition groups to represent resource competition relationships among operators within the same group. The edges' weights are self-adjusted by an attention mechanism that is coupled with the data flow structure. Next, we use a GAT model to calibrate operator costs. Finally, to gather those calibrated costs into the final QPP result, we develop a method to convert the graph back to a tree in a differentiable way and use a TCN to aggregate those costs from the bottom to the top and predict the query performance as \textbf{Stage 3}.

Our contributions can be summarized as follows:
\begin{enumerate}
    \item We designed a multi-stage query performance prediction approach that decouples operator cost estimation, resource contention calibration, and query performance prediction. This method provides more accurate query performance forecasting for database systems with complex underlying query mechanisms (such as high-performance OLAP), while also filling the research gap in learning-based cost estimation for database systems featuring dynamic parallel DAG pipelines.
    \item We design a low-level \textsf{Runtime Tracker} and implement it on ClickHouse to collect detailed and accurate data for the QPP model's training and prediction, which includes full-collection mode, probe execution mode, \textsf{Cost Logger}, and a modified serial executor.  
    \item Experimental results demonstrate that CONCERTO achieves higher estimation accuracy and comparable or lower time overhead than baseline methods.
\end{enumerate}

\section{Related Works}
\subsection{Conventional Methods}
The most common method is to build cost models for different operators before learning methods are introduced into the database system. Such methods employ the CPU cost, I/O cost, etc. of operators to estimate the query cost. Manegold et al. proposed a series of cost functions for different memory access patterns~\cite{CostModel1}. These functions are parameterized to fit different hardware features. The operators' cost can be inferred from the combination of those cost functions. Feilong Liu et al. proposed a memory I/O cost model~\cite{CostModel2} to identify good evaluation strategies for complex query plans with multiple hash-based equip-joins over memory-resident data. Wentao et al. proposed a cost model~\cite{UAQPP} that represents the query latency as a function of the selectivities of operators in the query plan as well as the constants that describe the cost of CPU and I/O. They solve the coefficient of the cost model from the observed query information. Wu et al.~\cite{wu2013predicting} proposed an offline profiling method to set the coefficients of a cost model for specific hardware and software conditions.

Such methods have two main problems, the first problem is that the parameters of the cost models are hard to adjust properly to maximize accuracy. Although there are some methods ~\cite{UAQPP, AdaptiveQPP} proposed to adjust or solve the parameters, it's not accurate enough for dynamic workload environments since the ground truth costs change with the hardware workload. The second problem is that some research ~\cite{HowGoodAreQueryOptimizers, End2End} show that the cardinality estimation results are crucial to those cost models, which is another harder problem that usually has much higher error. One important reason is that, compared to learned models, conventional methods' prediction results highly depend on the input scale and lack robustness.

\subsection{Learned Methods}
Before the rise of deep learning, some studies used machine learning to predict query performance. Ganapathi et al. take plan-level information into consideration and use KCCA~\cite{KCCA} to predict multiple performance metrics for a single query\cite{QPP-KCCA}. This method can support both short and long-time-running queries~\cite{Huang2023SurveyOP}. Li et al.~\cite{li2012robust} train a boosted regression tree for every operator and estimate the cost of the whole plan by scaling and adding the operators' cost. Akdere et al. proposed two QPP models~\cite{OpModelCombine}. The first is a plan-level model that takes the input of the query plan information, including numbers and cardinality of each operator, and predicts the overall query performance. The second is an operator-level model that predicts the operator's cost based on the feature of the operator itself and the cost of its child operators. Then they combined these two methods by using the plan-level model to deal with the materialization of the sub-query plan and the operator-level model for others, in order to avoid the high error of the operator-level models on the materialization sub-plan.

In the area of deep learning-based QPP methods, Marcus et al. proposed NEO~\cite{NEO} and QPPNet~\cite{QPPNet}. The NEO~\cite{NEO} captures relationships between physical operators and their child operators, they do not effectively capture resource competition among operators across different sub-trees. The QPPNet employs one deep learning-based latency model for each operator. Each model accepts the latency, data, and query features of its children operators and other features and then predicts the latency of its operator. Then it dynamically combines the networks corresponding to all operators based on the structure of the query plan tree and passes data features, query features, and latency information between operators to estimate the overall query latency. It's a flexible method that can adapt to various tree-structured query plans. However, it has a very high cost to build the network for each plan, and an insufficient training throughput due to the heterogeneous network architecture makes it hard to take full advantage of the GPU’s data parallel capabilities.

Ji Sun et al. proposed an end-to-end learned cost estimator~\cite{End2End} which includes carefully designed complete query representation and feature engineering by vectorizing encoded physical query plan trees containing predicate information and uses a tree-LSTM model to perform end-to-end prediction of query costs. This method is also designed for a tree-shaped physical plan. 

Yue Zhao et al. proposed QueryFormer~\cite{QueryFormer}, a variant of Transformer that can capture tree-shaped data modal more effectively, which can be used for tasks like QPP. 

Lin Ma et al. proposed MB2~\cite{MB2}, a two-level QPP method that considers concurrent operators. They first build a lightweight resource cost model for each operator, then use an interference model that takes inputs of the estimated cost vector of operators and predicts the query performance. Compared to previous methods, this method is more suitable for OLAP databases that involve parallel query plans. However, the interference model didn't use any special design to capture the resource competition between operators. 

Xuanhe et al. proposed a QPP method for concurrent queries using graph embedding~\cite{GraphEmbQPP}. They added edges for operators that have resource sharing or competition between query plans of concurrent queries to construct a DAG. Based on resource competition and data sharing, numerical features are added to the edges. Then a graph embedding model is employed to capture the information among all queries and predict the query performance.

Benjamin et al. proposed a zero-shot cost (latency) prediction method~\cite{Zero-shot} to achieve robustness on various data, workloads, and physical designs of databases by carefully designing query representation and using GNN to perform prediction. 

Li Yan et al. proposed RAAL~\cite{ResourceAwareQPP}, a resource-aware deep cost model, which proposed a node-level resource-aware attention model to predict query latency.

However, none of these studies considered the resource competition within a single parallel query plan and the dynamic modification of pipelines in high-performance OLAP databases.

\section{Overview}
\label{Section:Overview}
\begin{figure*}[htbp]
  \centering
  \includegraphics[width=0.9\linewidth]{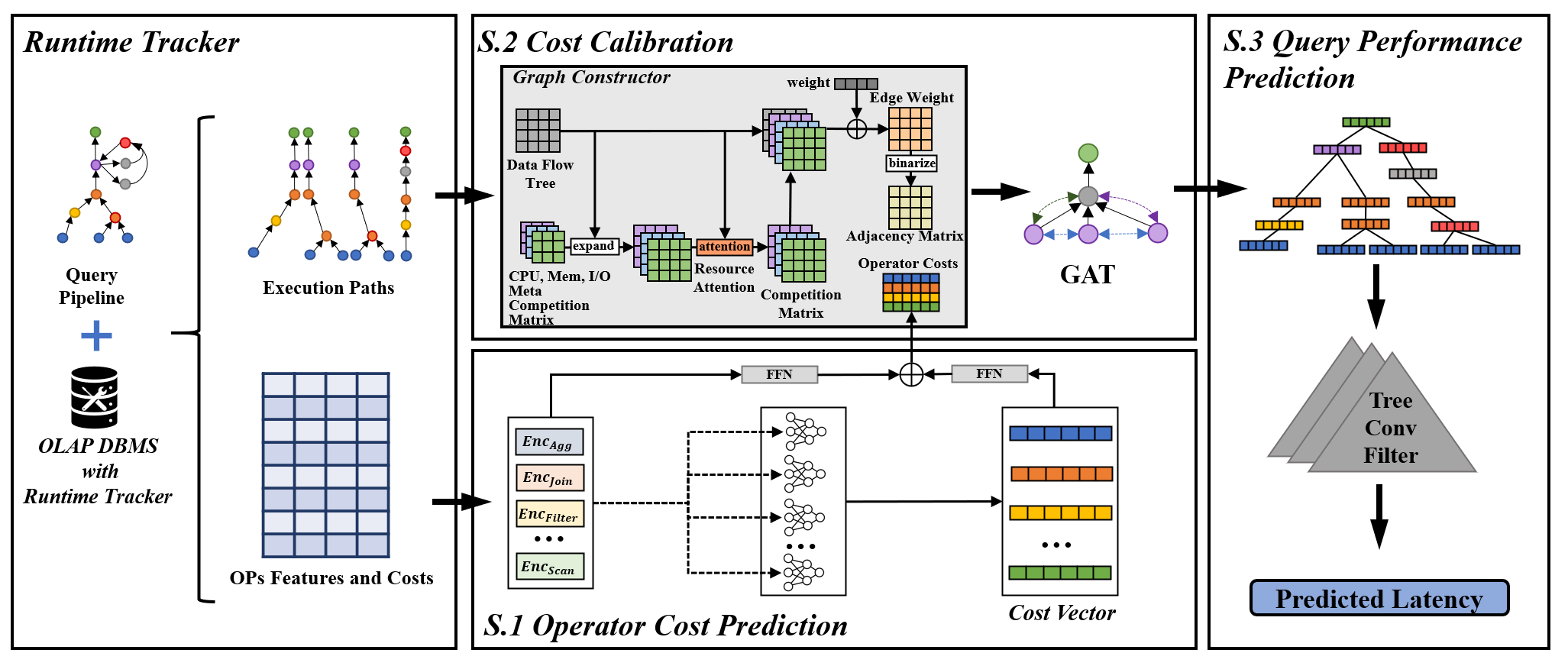}
  \caption{Overview of CONCERTO, which shows the process of data collection, training, and prediction.}
  \label{fig:overview}
\end{figure*}
As introduced in \autoref{section:Introduction}, CONCERTO decouples the cost prediction process of operators and the whole query. Besides the OCPs, such a workflow requires a method to collect detailed runtime resource costs, time-cost information, and operators' features to train those OCPs. It also includes a multi-stage model to summarize each operator's cost of the query plan and predict the total query latency. As shown in \autoref{fig:overview}, CONCERTO contains three stages: 
\begin{itemize}
    \item \textbf{Stage 1: Operator Cost Prediction}. As the \textbf{Stage 1}, it responds to estimate each SIMD operator's cost by the \textsf{OPCs}, a set of lightweight neural networks responsible for predicting the corresponding operators' resource costs. Each model is trained on the operator cost data collected by the \textsf{Runtime Tracker}. They take the system overhead and operators' calling information as input and output the predicted cost vector of operators. Note that the cost vector is not limited to latency only, but includes the hardware resource overhead that the operator is expected to consume without considering the impact of other concurrent operators. For a given database software and hardware environment, compared to end-to-end models, OCPs can provide better generalization performance across different query plans. Finally, the cost vectors and vertex features are mapped by FFN separately and added together for later cost calibration.
    
    \item \textbf{Stage 2: Cost Calibration.} In \textbf{Stage 2}, the \textsf{Cost Calibrator} first uses the \textsf{Graph Constructor} to merge execution paths collected in the probe execution mode of the \textsf{Runtime Tracker} into a data-flow tree as a rich expression of the query plan. Then it uses an attention mechanism to adjust each operator's resource competition weight based on both the meta-competition matrix and the data-flow tree. After that, the \textsf{Cost Calibrator} uses a GAT to calibrate each operator's cost vector based on the DAG and predicted operators' costs during parallel pipeline execution.
    
    \item \textbf{Stage 3: Query Performance Prediction.} In \textbf{Stage 3}, we combine the calibrated cost vectors and the structure of the data-flow tree into a vector tree and use a TCN model to summarize every operator's cost vector from the bottom up and predict the total latency of the whole query. It should be noted that the entire process during Stages 2 and 3 is differentiable, thus, we can pass the gradient back to the GAT network and train the GAT and TCN models together. 
\end{itemize}

To support the three stages above, CONCERTO also includes a \textsf{Runtime Tracker}, which can collect necessary data for training and inference. The \textsf{Runtime Tracker} made some modifications to the common components like the physical operators, data chunks, and executors. We will introduce it in detail in \autoref{Section:CostTracker}.

In the aspect of the training phase, the \textsf{Runtime Tracker} first collects each operator's features and performance information by executing queries with a serial executor under the full-collection mode, which is either provided by the OLAP DBMS or modified from the OLAP DBMS's parallel executor by the \textsf{Runtime Tracker}. Then it collects the query plan and the whole query's latency with the DBMS's parallel executor under the probe execution mode. Next, we use the collected training data to train the OCPs. The last step of training is to use the predicted cost vectors and other runtime pipeline data to train the \textsf{Cost Calibrator} and the Query Performance Predictor together. 

In the aspect of the prediction phase, we only need to set the \textsf{Runtime Tracker} to probe execution mode, which uses the DBMS's parallel executor for a fast probe execution to obtain each operator's runtime information, the chunks' execution paths, and the system overhead. Then we use the three stages, i.e., Operator Cost Prediction, Cost Calibration, and Query Performance Prediction, sequentially to predict the query's latency.

In the rest of this section, we will introduce CONCERTO's \textsf{Runtime Tracker}, and the three stages will be introduced in \autoref{Section:Methods}.

\section{CONCERTO's QPP Model}
\label{Section:Methods}

In this section, we introduce CONCERTO's predictor part in the following order: Operator Cost Prediction, Cost Calibration, and Query Performance Prediction.

\subsection{Operator Cost Prediction}
\label{subSection:OperatorCostPrediction}

In this section, we introduce the Operator Cost Predictor (OCP) in detail. To predict high-performance OLAP DBMSs' query performance, it is necessary to estimate the resource consumption and the current system resource utilization. Knowing that each operator's resource consumption is the precondition for predicting the resource competition between operators and accurate latency prediction, since those DBMSs are designed to maximize resource utilization efficiency with a high degree of parallelism.

The OCPs are designed to predict the resource consumption of the given operator without the interference of other concurrent operators. Therefore, we can decouple and predict the cost of a single operator and the cost of the whole parallel query pipeline separately. This can improve the prediction accuracy for physical operators with SIMD instructions and make it easier for CONCERTO to generalize between different query templates to reduce the accuracy degradation problem caused by workload drift. The inputs of the OCPs consist of two parts: Resource Utilization and Operator Features.

\begin{itemize}
    \item Resource Utilization: It contains CPU, memory, and I/O utilization before execution; all OCPs have these three features.
    \item Operator Features: The operator features include the operator type, calling parameters, and data features. For each operator, we select the parameters that have the most significant influence on their cost as a part of the feature. Some of them are determined by knobs, and some are determined by the executor at runtime. The data feature is provided with the input rows (cardinality) and columns. The information on whether an operator is implemented by the SIMD instruction set will be captured by the OCP model. This type of feature varies depending on the operators and the compile settings.
\end{itemize}

There are two types of features, and we encode them separately. For numerical features, we normalize them during encoding to avoid numerical issues. For categorical features, including string-type features and other non-numerical features, we use one-hot encoding to represent them.


Since the goal of OCPs is to predict each operator's resource cost, their output is defined as the cost vector that includes the cost of CPU, memory, and I/O. The CPU cost includes the elapsed time, CPU time, CPU cycles, CPU instruction numbers, CPU cache references, and CPU cache misses. The memory cost includes the average and maximum memory consumption. The I/O cost includes the disk block read and write numbers. We use shallow MLP networks as OCPs to achieve a good balance between accuracy and inference speed and train them with the MSE loss function.

\subsection{Cost Calibration}
\label{subSection:CostCalibration}

Cost calibration constitutes a critical component of CONCERTO, particularly for high-performance OLAP DBMSs that extensively employ parallel execution paradigms. These systems typically incorporate three fundamental parallelism strategies: (1) vectorized operator implementations, (2) intra-operator multithreading, and (3) parallel pipeline execution. 

CONCERTO's decoupled architecture handles SIMD-optimized operators through dedicated Operator Cost Predictors (OCPs). For intra-operator parallelism, the pipeline structure explicitly models multithreaded execution, with cost calibration performed by our proposed Graph Attention Network (GAT) model. The most challenging scenario arises from vertical parallel execution, where the DBMS continuously traverses the pipeline and concurrently executes all data-ready operators. While this design maximizes throughput, it introduces system-wide contention for computational resources (CPU cycles), memory bandwidth, and I/O operations across the entire pipeline.


To calibrate those operators' costs predicted by the OCPs, CONCERTO needs to know the pipeline structure and the resource competition relation among all operators. As we discussed in \autoref{Section:CostTracker}, the regular `EXPLAIN' tool is not enough for pipelines with dynamic modification mechanisms, and CONCERTO's cost tracker collects more detailed information, including tracked execution paths and other physical plan information, with probe execution mode. In stage 2 of cost calibration, CONCERTO's \textsf{Graph Constructor} fully uses the tracked execution paths of chunks and constructs a data-flow tree instead of using the DAG pipeline.

To understand such data-flow tree construction better, taking the \autoref{fig:DynamicModification} as an example, it shows the pipeline and chunks during the probe execution. Assuming that the high-performance OLAP DBMS dynamically added and replaced some operators, such as MergeSort, during the probe execution\footnote{\autoref{footnote:pipeandDAG}}, then all chunks after this modification would be executed through a different path. Suppose that we simply use the modified DAG pipeline as a representation of the query plan. In that case, all execution information before modification has to be dropped since it cannot match the modified pipeline. Such a solution will cause inaccurate estimation. Besides, such a DAG pipeline is also hard to adapt to existing methods for comparison and hard to support other downstream tasks since most of them are designed based on a tree plan that is used by most DBMSs. Therefore, CONCERTO takes full use of the information collected during the probe execution mode by merging each chunk's execution path into a data-flow tree, in which different pipelines exist at the same time. 

As shown in \autoref{Algorithm:ConstructDataFlowTree}, we locate each operator by a chunk's execution path that starts at the root operator and ends at itself. We first traverse every execution path and split the path into a series of operators' IDs at lines 2-3. Then we traverse each operator on that path from top to bottom at line 4. We first check whether the current operator is the root node of the data-flow tree and set it at lines 5-7. If not, then we try to match it in the children of the current node from line 8. If the current node does not have any children, we create a new node for the current operator and connect it as a child of the current node at lines 9-12. Otherwise, we try to match each child node with the current operator at lines 13-15. If we get a match, then we set the matched child node as the current node and start the next round of the loop to match the next operator at line 16. Otherwise, we also create a node with the current operator and connect it as a child at lines 17-20.

By applying such a construction algorithm, the data-flow tree will branch at the red part of the pipeline in \autoref{fig:DynamicModification} since the modified execution path will not match at line 15 of \autoref{Algorithm:ConstructDataFlowTree}. Therefore, the data flow during probe execution is expanded from the time dimension to the spatial dimension, which can express pipelines before and after modification. 

\begin{algorithm}[htbp]
    \footnotesize
    \caption{Data-flow tree construction}
    \label{Algorithm:ConstructDataFlowTree}
    
    root\_node $\leftarrow$ null;\\
    \ForEach{chunk\_path\ in\ chunk\_paths}{
        operator\_list $\leftarrow$ spilt(chunk\_path);\\
        \ForEach{operator\ in\ operator\_list}{
            \uIf{operator\ is\ root}{
                node $\leftarrow$ CreateNode(operator);\\
                root\_node $\leftarrow$ node;
            }
            \uElse{
                \uIf{node.children.size==0}{
                    new\_node $\leftarrow$ CreateNode(operator);\\
                    node.children.append(new\_node);\\
                    node $\leftarrow$ new\_node;\\
                }
                \uElse{
                    \ForEach{child\_node in node.children}{
                        \uIf{match\_operator\_in\_the\_same\_layer(child\_node, operator)}{
                            node $\leftarrow$ child\_node;\\
                        }
                        \uElse{
                            new\_node $\leftarrow$ CreateNode(child\_operator);\\
                            node.children.append(new\_node);\\
                            node $\leftarrow$ new\_node;\\
                        }
                    }
                }
            }
        }
    }
    \Return root\_node;
\end{algorithm}

\begin{figure}[htbp]
  \centering
  \includegraphics[width=\linewidth]{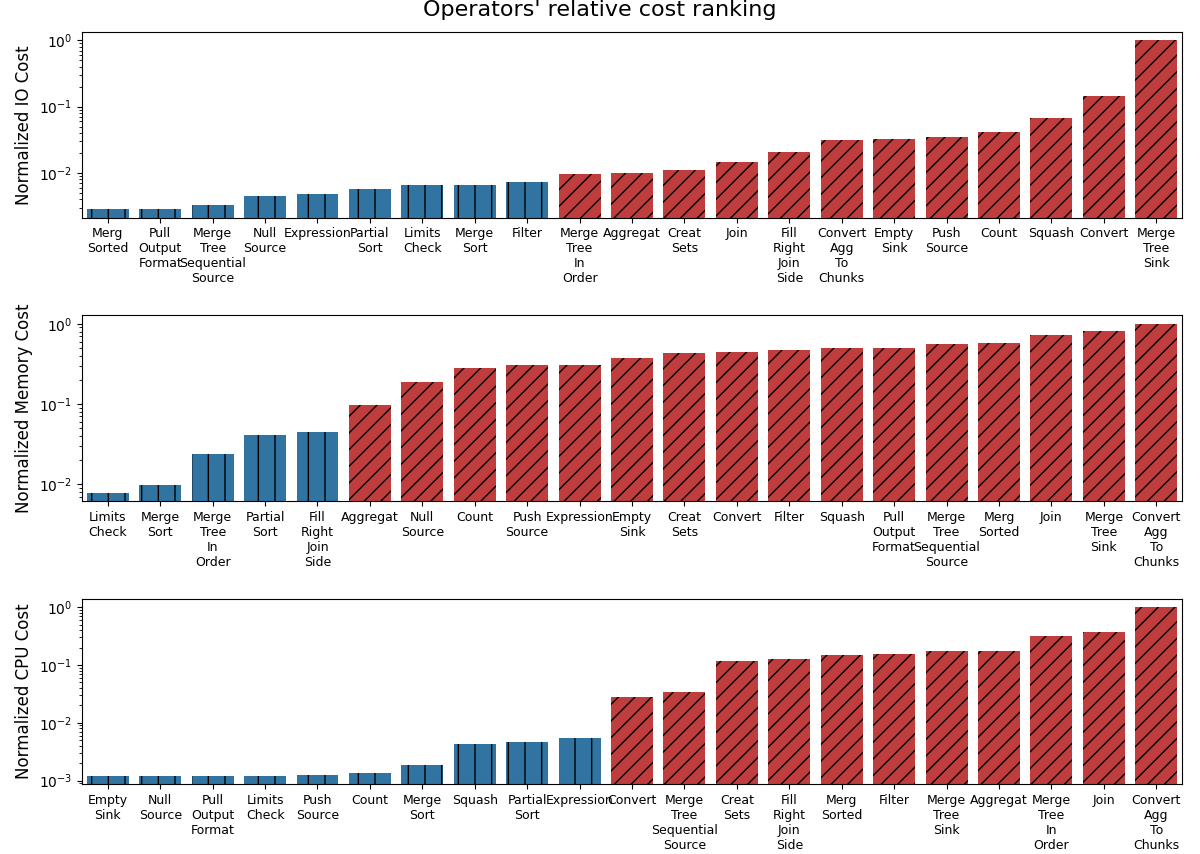}
  \caption{Illustration of ClickHouse's physical operators' relative resource cost. The red operators are marked as existing corresponding resource competition.}
  \vspace{-5pt}
  \label{fig:OperatorResourceCompetition}
\end{figure}

The data-flow tree is used as the backbone of the input graph of the \textsf{Cost Calibrator}, where edges representing resource competition are added. Now we explain how the \textsf{Graph Constructor} generates the CPU, memory, and I/O resource competition. We evaluated each operator's resource consumption to divide them into different resource-bounded types based on both code implementation and the number of their CPU instructions, memory cost, and I/O times while processing the same amount of data. As the results are shown in ~\autoref{fig:OperatorResourceCompetition}, an operator can have multiple types of bounded resources. Based on this, the \textsf{Graph Constructor} can build CPU, memory, and I/O meta-resource competition matrices between every distinct operator type separately. Let the number of operator types be $M$. Those matrices can be denoted as $\mathcal{M}^c$, $\mathcal{M}^m$, $\mathcal{M}^{io} \in R^{M \times M}$. Note that this is only a rough classification, and the resource competition intensity will be adaptively learned based on the pipeline structure with an attention mechanism that will be introduced in the next paragraph. For a given data-flow tree $F$ constructed from a pipeline that has $N$ operators, CONCERTO treats it as a DAG and expresses it as an adjacency matrix $\mathcal{M}_{F} \in R^{N \times N}$. The \textsf{Graph Constructor} expands those meta resource competition matrices to $\mathcal{M}_{F}^c, \mathcal{M}_{F}^m, \mathcal{M}_{F}^{io} \in R^{N \times N}$ with the same dimension by indexing the operator's resource competition relation by the operators' order in $\mathcal{M}_{F}$. After being adjusted by an attention mechanism, the results $\mathcal{M'}_{F}^c$, $\mathcal{M'}_{F}^m$, $\mathcal{M'}_{F}^{io}$ are added to the $\mathcal{M}_{F}$ to obtain the final expression of the pipeline:
\begingroup
\begin{align}
\mathcal{M}_{pipeline} = \mathcal{W}_{F} \mathcal{M}_{F} + \mathcal{W}_{c} \mathcal{M'}_{F}^c + \mathcal{W}_{m} \mathcal{M'}_{F}^m + \mathcal{W}_{io} \mathcal{M'}_{F}^{io} \label{formula:M_pipeline}
\end{align}
\endgroup
, where $\mathcal{W}_{F}, \mathcal{W}_{c}, \mathcal{W}_{m}, \mathcal{W}_{io}$ are learnable weights.

\textbf{Resource Competition Attention.} Although the resource competition matrices are summarized based on both operators' code analysis results and relative cost evaluation, resource contention between operators with resource competition relations may not occur during the pipeline execution process. Even if it does occur, the intensity of resource contention may vary depending on system load, pipeline structure, data characteristics, operator call parameters, etc. The existing work~\cite{GraphEmbQPP} uses estimated operators' latency to build a timeline of operators' calling and estimate the competition intensity by the operators' overlapped ratio. However, such a solution is vulnerable, and the error would accumulate with the growth of the operator's number since each operator's start and end time depend on every operator's latency before it. For high-performance OLAP DBMSs, the pipeline is usually longer and involves parallel branches. Besides, they execute operators immediately once the operator is ready to be executed. So when we dive to the level of each step, the execution time and order of operators are more random, so it is impossible to accurately predict the timeline and overlap the time between operators like in the existing work~\cite{GraphEmbQPP}. Instead, we introduce an attention mechanism to self-adjust operator competition intensity based on the given data-flow tree and resource competition DAG. 

The core idea of attention is to calculate the attention score that represents the correlation between every item in features $Q, K$ and apply that score to the feature $V$~\cite{AttentionIsAllYouNeed}. 

\begingroup
\setlength{\abovedisplayskip}{2pt}
\setlength{\belowdisplayskip}{2pt}
\begin{align}
Attention(Q,K,V) = softmax(\frac{QK^{\mathsf{T}}}{\sqrt{d_k}})V
\end{align}
\endgroup

In this case, CONCERTO needs to calculate the operator's resource competition intensity based on the given data-flow tree $\mathcal{M}_{F}$. Thus, the attention score should be calculated from the resource competition matrix $\mathcal{M}^{resource}_{F}$ and the data-flow tree $\mathcal{M}_{F}$, which can be formalized as:

\begingroup
\setlength{\abovedisplayskip}{2pt}
\setlength{\belowdisplayskip}{2pt}
\begin{align}
    \mathcal{M'}^{resource}_{F} = softmax(\frac{\mathcal{M}^{resource}_{F}\mathcal{M}_{F}^T}{\sqrt{N}})\mathcal{M}^{resource}_{F}
\end{align}
\endgroup

, where $N$ is the number of involved operators of the pipeline, $resource \in \{c, m, io \}$, and $\mathcal{M}^{resource}_{F},\mathcal{M}_{F}^{\mathsf{T}} \in R^{N \times N}$. The final resource competition intensity that is fed into the downstream model is summarized by \autoref{formula:M_pipeline}:


After constructing the graph, CONCERTO uses a GAT model~\cite{GAT} to capture the resource competition and calibrate the operators' cost vectors. 


After being processed using all the methods above, CONCERTO now gets the calibrated resource cost vector of each operator in the pipeline. CONCERTO then converts the calibrated resource vectors back to the pipeline tree structure and summarizes those vectors from the bottom to the top by a TCN to predict the query performance. Note that all learnable weights in this section are trained by QPP tasks together with TCN introduced in the next subsection.

\subsection{Query Performance Prediction}
In this section, we introduce how CONCERTO predicts the query performance based on calibrated cost vectors and how to train the calibrator and the predictor together.

In \autoref{subSection:CostCalibration}, we express the pipeline as a data-flow tree, and each operator's predicted cost vector is calibrated according to the DBMS's operator parallelization mechanism. Considering the high-performance OLAP DBMSs' execution mechanisms, simply adding each operator's cost vector as the final predicted performance is too rough. Instead, CONCERTO combines those calibrated cost vectors according to the data-flow tree structure to construct a vector tree. We use a tree convolutional network (TCN)~\cite{TCN} to summarize the cost from bottom to top. The TCN has two main advantages. Firstly, it is powerful to capture tree shape query plan's feature with a low inference cost, which is the reason why it has been widely used in many AI4DB methods~\cite{NEO, ASurvey}. Secondly, the paper of TCN proposed a \textit{continuous binary tree} mode, which natively supports trees with multiple child nodes. Since the data-flow tree is constructed from execution paths of a DAG pipeline, many nodes have more than two children, and TCN can handle such a scenario well.

The TCN has a triangle-shaped convolution filter that is made of three weight matrices $W^t_{conv}, W^l_{conv}, W^r_{conv}$. For node $x_i$ in the window of a filter, its weight matrix is denoted as $W_{conv, i}$, which is a linear combination of those three weights above with corresponding coefficients $\eta_i^t,\eta_i^r$, and $\eta_i^l$. Each layer's output of TCN is calculated by the following formulas~\cite{TCN}:

\begingroup
\setlength{\abovedisplayskip}{2pt}
\setlength{\belowdisplayskip}{5pt}
\begin{align}
    \eta_i^t = \frac{d_i-1}{d-1} \quad \eta_i^r &= (1-\eta_i^t)\frac{p_i-1}{n-1}, \quad \eta_i^l = (1-\eta_i^t)(1-\eta_i^r)\\
    W_{conv, i} &= \eta_i^t W^t_{conv} + \eta_i^l W^l_{conv} + \eta_i^r W^r_{conv}\\
    y &= tanh(\sum^n_{i=1}{W_{conv, i} \cdot x_i + b_{conv}}) 
\end{align}
\endgroup
, where $d_i$ is the depth of node $i$ in the window, and $d$ is the depth of the window.

CONCERTO's GAT and TCN networks are trained together in a query-driven way since constructing the calibrated cost vector to the cost vector tree is differentiable. They are trained with the MSE loss function together with normalized ground-truth query latency as the label. To make the model converge more stably, all neural networks are connected by a residual connection path~\cite{ResNet}.
\label{subSection:QueryPerformancePrediction}

\section{Runtime Tracker}
\label{Section:CostTracker}

\subsection{Motivation}
To develop a physical plan-level performance predictor for high-performance OLAP DBMSs, we need to collect detailed information to train and evaluate the predictor. However, such DBMSs' parallel and dynamic execution mechanisms prevent us from using tools like `EXPLAIN' to support query performance prediction for the following reasons.

\begin{figure}[htbp]
  \centering
  \includegraphics[width=0.6\linewidth]{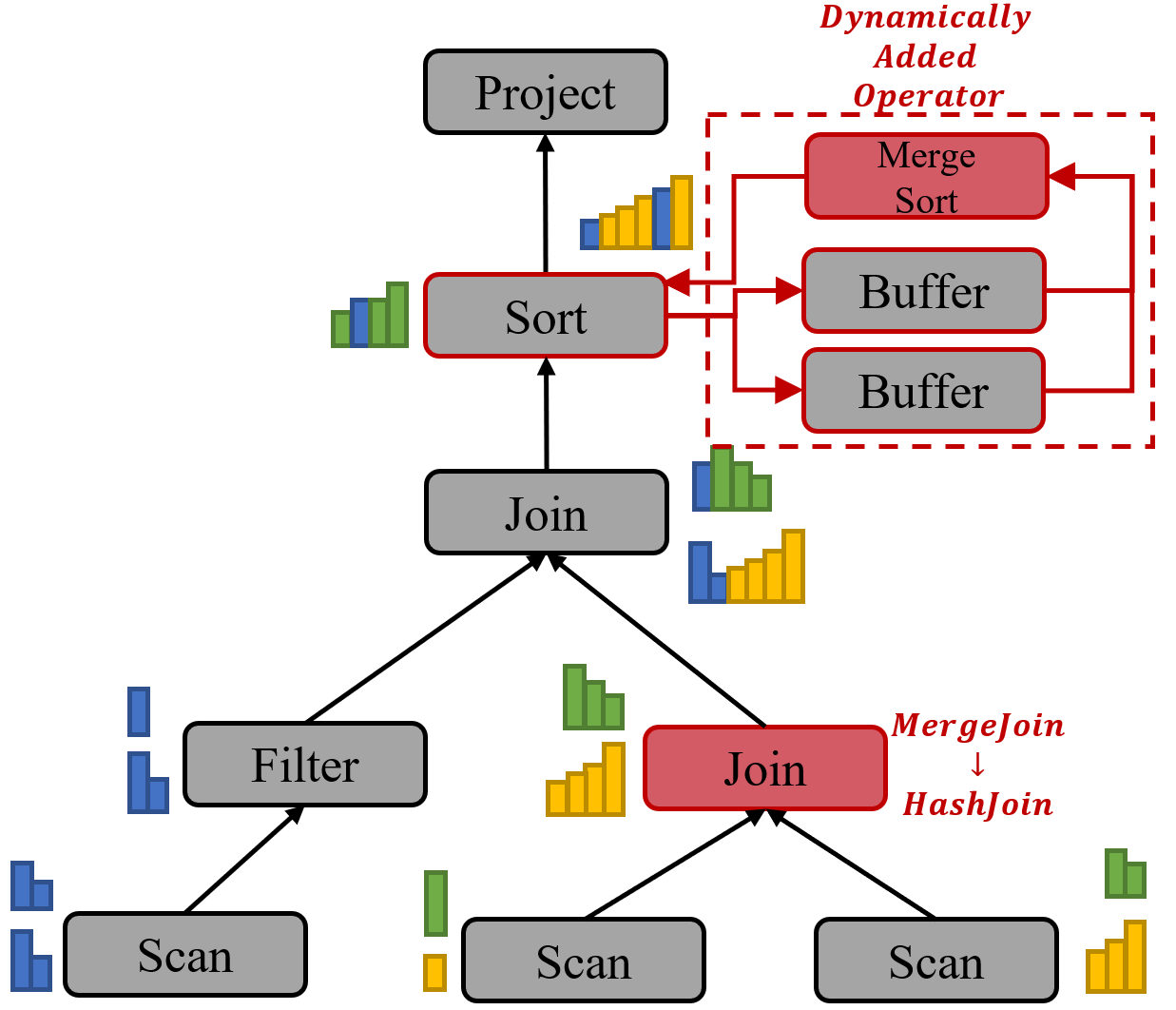}
  \caption{Illustration of dynamic modification. The Join operator with a red background is modified from hash join to merge join and the MergeSort operators are added to the DAG in the probe execution phase.}
  \label{fig:DynamicModification}
\end{figure}

To make an accurate query performance prediction, it is necessary to acquire detailed operators' runtime information, such as their resource cost and runtime calling parameters. It would be hard to match each operator's information with the pipeline structure given by the `EXPLAIN' SQL command since most of them employ dynamic execution mechanisms that allow inserting and modifying operators including adding new execution paths and changing the algorithm used by operators such as \textit{sort} and \textit{join}, which is hard to foresee during performance prediction. 

\autoref{fig:DynamicModification} shows a typical dynamic pipeline modification scenario if a DBMS with a dynamic execution mechanism (a.k.a adaptive query execution in SparkSQL and Databricks) finds that the pipeline is not the optimal one under the current system workload after launching the query, it will dynamically modify the pipeline by two types of actions: (1) changing the algorithm used by the corresponding operator; (2) changing the structure of the pipeline by inserting new operators.

For the first type of action, assuming the DBMS finds that it mistakenly selected the merge join instead of the hash join algorithm, it will actively modify the join algorithm. Some DBMSs~\cite{clickhouse} achieve this by passing different types of algorithm parameters to the \textit{Join} operator; the instance of the operator remains the same, but the operator will have different features and costs for most of the execution, which cannot be obtained by simply using the static pipeline. Other DBMSs, such as SparkSQL, change the algorithm by replacing the corresponding physical operator. In both situations, relying solely on the information provided in a static pipeline is unreliable. 

For the second type of action, assuming that the DBMS decides to switch from internal sorting to external sorting, it will involve adding and replacing operators in the pipeline. Some DBMSs, like SparkSQL with the AQE mechanism, can even adjust the join order during execution. In such a situation, the execution paths of chunks can be completely changed. 

Therefore, simply extracting static information from the planner is not enough; the runtime information, including calling parameters, pipeline modifications, and chunks execution information, is necessary for precise performance prediction. However, the pipeline exported by the `EXPLAIN' statement not only does not contain any of those modification details, but it also cannot provide any runtime information about operators and chunks. Thus, it is difficult to use for QPP. Although it is impossible to obtain all pipeline modifications during the whole execution process while predicting query performance, inspired by ClickHouse's probe phase~\cite{clickhouse}, we developed a probe execution mode to launch, track, and collect the first several chunks' execution information rapidly and obtain the pipeline structure, operators' runtime features and pipeline modifications happened on the starting of the execution. 

Considering the dynamic pipeline execution mechanism, those chunks can be generated by different execution paths that correspond to different features and costs. Thus, it would be a time-space graph with unaligned time steps. To collect that information, CONCERTO either embeds a variable number of chunks' features from unaligned time slices into each operator of the pipeline or separates those chunks' execution paths by constructing and processing a data flow tree. The former solution needs time-space graph convolution on a graph with many padded zero features, which could lead to resource waste and lower accuracy. Therefore, we chose the latter solution, which expands the execution path of multiple chunks involved in the probe execution in the spatial dimension and constructs a data-flow tree by combining them as the representation of the dynamic pipeline structure. 

Besides, since CONCERTO is a multi-stage method that decouples the operator cost prediction, cost calibration, and query performance prediction, it also has to collect the training data of OCPs separately to avoid interference from other parallel operators.

With all those reasons above, we designed the \textsf{Runtime Tracker} to collect detailed runtime information to support more accurate QPP. For the convenience of discussion, we take ClickHouse, a representative high-performance OLAP DBMS, as an example to introduce the \textsf{Runtime Tracker}'s design in the following subsection. Note that such a design is portable between DBMSs since high-performance OLAP DBMSs have similar execution mechanisms such as shown in \autoref{tab:summary}, and we assume the target DBMS uses a DAG-based pipeline, which is a more general structure compared with the tree-shaped pipeline.

\subsection{Design of \textsf{Runtime Tracker} and Data Collection Process}

\begin{figure}[htbp]
  \centering
  \includegraphics[width=0.9\linewidth]{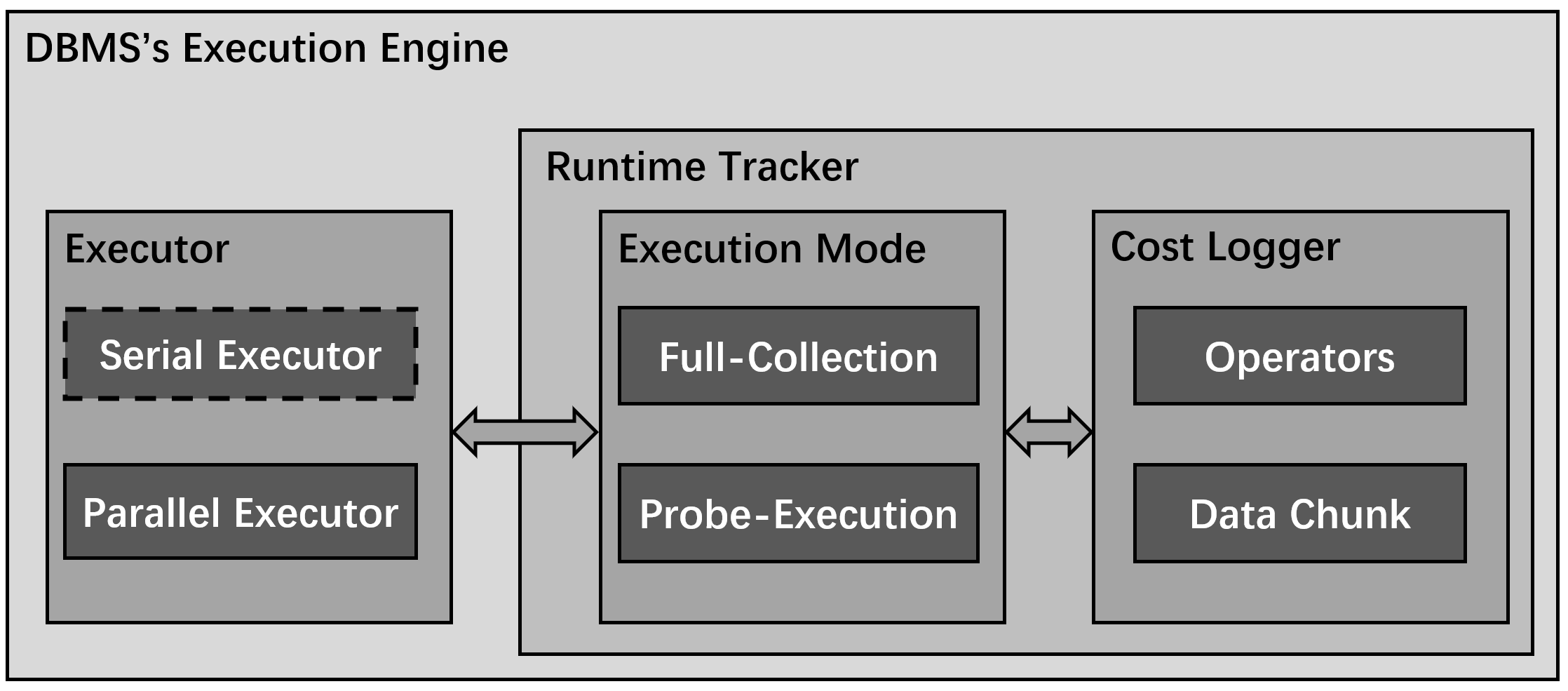}
  \caption{Illustration of the structure of the \textsf{Runtime Tracker}. The Serial Executor in dotted lines needs to be modified from the DBMS's parallel executor if the DBMS does not provide a serial execution mode like SparkSQL does.}
  \label{fig:RuntimeTracker}
\end{figure}

As shown in \autoref{fig:RuntimeTracker}, the \textsf{Runtime Tracker} includes three parts: 

\begin{itemize}
    \item \textbf{Executor.} The executor part can usually be regarded as a part of the DBMS's execution engine. However, in order to obtain the OCPs' training data without interference from the parallel execution of other operators, CONCERTO needs a serial executor. Some DBMSs like SparkSQL provide such a serial execution mode, but some like ClickHouse do not. For such DBMSs, it is necessary to create a serial executor from their parallel executor. Such modification is usually simple since most DBMSs with parallel executors provide a knob for the degree of parallelism (DOP).
    
    \item \textbf{Execution Mode.} CONCERTO uses a full-collection mode to collect the training data for Stage 2 and Stage 3. Since CONCERTO collects the chunk-level information that cannot be fully obtained during inference, it employs a probe execution mode to obtain a part of the dynamically changing pipeline information.
    
    \item \textbf{\textsf{Cost Logger.}} CONCERTO uses very detailed low-level information about the physical pipeline as features, including the operator's cost, calling parameters, system resource utilization, and chunk execution information. The information is finally collected by \textsf{Cost Logger}, which involves some simple wrappers and modifications of the DBMS's physical operators and data chunk class.
\end{itemize}
\vspace{-3pt}
To explain how the \textsf{Runtime Tracker} collects data, we introduce it from the perspective of data collection for different parts of the models. 

\textbf{Data for OCPs.} The features and execution costs of each single operator that is used to train the OCPs should be collected offline. Since the OCPs aim to predict the cost of each operator without interference from the parallel execution of other operators, their training data should be collected separately by a serial executor. To collect enough data with as few queries as possible, we run \textsf{Runtime Tracker} in full-collection mode, which collects every operator's runtime features and costs on every data chunk. Although this will cause extra data collecting and training costs compared to using the history workload log, the overhead would not be too high because the OCPs are small MLP networks, and they only need a single-time data collection for a deployed DBMS. 

We wrap the \textit{work()} function of operators with our \textsf{Cost Logger} to record the operator's calling parameters, system resource utilization, and resource cost during execution. The \textsf{Cost Logger} uses \textit{ioctl} library to obtain the CPU instructions, CPU cycles, cache references, and cache misses. It uses the \textit{cpuinfo} file to obtain the CPU frequency and uses \textit{proc/stat} to obtain the CPU utilization. The \textsf{Cost Logger} also uses the \textit{proc/meminfo} file of Linux to measure the memory cost and uses the \textit{/proc/PID/io} file to measure the I/O read and write cost. After applying these modifications, the serial executor can be regarded as degenerating into a volcanic model. Then we execute workloads and collect each operator's features and the costs of executing each data chunk. The data collection is efficient because there are dozens to hundreds of data chunks for each operator in one query.
    
\textbf{Data for \textsf{Cost Calibrator}.} As discussed above, CONCERTO solves the dynamic modification of pipeline and calling parameters during cost calibration and query performance prediction by expanding the runtime execution paths in the spatial dimension. This is achieved by tracking the execution path of chunks in the probe execution mode. During offline training data collection, CONCERTO will set the executor and the tracker to the probe execution mode and make a quick, short execution till the first few result chunks are collected. During this procedure, all intermediate chunks' execution paths and operators' calls are traced, and the potential dynamic modifications are also logged. The \textsf{Graph Constructor} will use this information in Stage 2 to construct a data-flow tree.


To track and log this information, we modified the code of the chunk and operators. We use the address of each operator instance as its ID to distinguish them. Once chunks are generated or processed, the corresponding operator will append its ID to the \textit{transform\_addr} list of the chunks. During the processing of chunks, \textsf{Cost Logger} would trace the costs and log them. The address of those cost records will also be added to the chunks' \textit{record\_addr} list. By analyzing those records, CONCERTO can construct the data-flow tree for cost calibration and query performance prediction. The detailed construction algorithm will be introduced in ~\autoref{subSection:CostCalibration}.

Taking the TPC-DS workload as an example, the probe execution phase's overhead is about 155ms. With this extra cost, we can obtain the data flow and partial dynamic added execution paths to support more accurate performance prediction. Since high-performance OLAP DBMSs are designed for processing complex OLAP queries that take hours to execute, the overhead is relatively small. Besides, this overhead is eliminable by marking those tuples that had been processed during probe execution and reading the result to skip them during the latter query execution.  

This approach appears to have a limitation in its inability to capture all dynamic modifications occurring throughout the entirety of query execution, particularly those stemming from system load variations, since they are unpredictable. However, it remains effective in detecting modifications arising from estimation errors, such as cardinality errors. These types of adjustments typically occur during the early stages of query execution, which are adequately addressed by the probe phase.

\section{Experiments}
In this section, we evaluate CONCERTO's performance and compare it with other baseline methods. Since ClickHouse supports all the mechanisms as shown in \autoref{tab:summary}, we consider it one of the most representative high-performance OLAP DBMSs. Thus, we implement CONCERTO on ClickHouse for evaluation.

\subsection{Experimental Settings}

\textbf{Baselines.} Since previous works already applied full comparison with conventional methods such as SVM and RBF~\cite{QPPNet, SVMQPP, RBFQPP} and proved learned QPP approaches outperformed conventional approaches on average accuracy, we will focus on the deep learning-based methods proposed in recent years. All baselines compared are listed as follows:

\begin{enumerate}
    \item QPPNet~\cite{QPPNet}: The QPPNet is an operator-level method that builds a cost model with a unified input and output interface for each operator and combines them according to the structure of the query plan tree to predict the latency. This method is considered to be a flexible model that can adapt to all different plan templates.
    \item QueryFormer~\cite{QueryFormer}: The QueryFormer is a variant of the Transformer, which provides an attention mechanism on tree-shaped data and can capture a high-level expression of query plans to support downstream tasks such as QPP.
    \item RAAL~\cite{ResourceAwareQPP}: The RAAL uses an attention-based resource-aware network to capture the resource allocation in SparkSQL and predict the query latency. The application scenario is quite similar to high-performance OLAP DBMSs' parallel pipeline, so we include it as a competitive baseline. We implement it by following the formula and instructions in its paper.
    \item GCN ~\cite{GCN}: The GCN is a well-known network that is proposed to perform convolution on graph data. Since both DAG and tree-shaped pipelines can be regarded as a graph, we use GCN to perform end-to-end QPP as a baseline.
    \item TCN ~\cite{TCN}: The TCN is a common and widely used network in the domain related to query execution and query optimization~\cite{NEO}, which introduces a convolution operation to the tree-shaped data. Thus, we included it as a baseline.
\end{enumerate}

We used their open-source code and modified it to adapt the data collected by CONCERTO's \textsf{Runtime Tracker} from ClickHouse due to ClickHouse's lack of a tool to export a detailed enough physical pipeline for baselines such as QPPNet and QueryFormer. In fact, CONCERTO's \textsf{Runtime Tracker} provides even more comprehensive features than more DBMSs' original implementation. For RAAL, since it is an approach designed for SparkSQL that considers the resource competition, which is also included in CONCERTO's scenario, it can share the same training and testing data with CONCERTO (only in a different format). Thus, the comparison is carefully designed and implemented to ensure fairness.

In addition to those baselines above, we also performed ablation experiments on CONCERTO to evaluate each component's performance with the following setups:
\begin{enumerate}
    \item CONCERTO w/o ResAttn: This combination only removes the resource attention mechanism of CONCERTO and uses the expanded meta-competition matrix only as GAT's input.
    \item CONCERTO w/o OCP: This combination removes the \textsf{OCP} module only and keeps the other modules to evaluate the impact of the \textsf{OCP} module on the QPP performance.
\end{enumerate}

\textbf{Datasets.} Followed by QPPNet~\cite{QPPNet} and MB2~\cite{MB2}, we compare the performance on TPC-H and TPC-DS benchmarks with $scale\_factor=1$ . These are two standard benchmarks designed specifically for evaluating OLAP database performance and are widely used in the industry.

\textbf{Workloads.} We use templates from TPC-H and TPC-DS for training and testing. The number of generated queries of each workload is about 10,000 (slightly fluctuating based on the number of query templates because we generate the same number of queries for each template). Among those queries, we evenly draw all the queries of 4 templates (templates 3, 7, 11, 6) from TPC-H (23.5\%) and 9 templates (templates 3, 9, 15, 21, 27, 33, 39, 45, 51, 57) from TPC-DS (15.52\%) by the index \textbf{as test workloads that will not appear in the training workloads}. Note that the testing workloads only use the data collected by the probe execution mode as input. 

\textbf{Metric.} We use the Q-Error as the metric to measure the accuracy of QPP, which is widely used in many previous QPP and cardinality estimation methods~\cite{QPPNet, AreOCMUnusable, Naru}. The definition of Q-Error is listed as follows: 

\begingroup
\setlength{\abovedisplayskip}{2pt}
\setlength{\belowdisplayskip}{2pt}
\begin{align}
    \text{Q-Error} &= \frac{\max{(prediction, actual)}}{\min{(prediction, actual)}} \label{metric:Q-Error}
\end{align}
\endgroup

\textbf{Model Architecture.} Considering the inference speed, CONCERTO uses relatively shallow networks, such as a 2-layer GAT and a 3-layer TCN, and all OCPs are MLPs with 3 hidden layers. 

\textbf{Hardware.} All training and testing workloads are generated on AMD R9-7945HX, 32GB memory, and RTX4060 with a controllable environment that has no interference from other processes. We use a public server with 6 RTX A6000 (48GB memory) GPUs, 2 Intel(R) Xeon(R) Silver 4210R CPUs, and 504GB RAM to perform all evaluations on CONCERTO, baselines, and ablation experiments.

\vspace{-10pt}
\subsection{QPP Accuracy}

\begin{table*}[!t]
\centering
\caption{Accuracy comparison of baselines, CONCERTO, and ablation studies.}
\vspace{-5pt}
\label{tab:accuracy}
\begin{tabular}{l|llllll|llllll}
\hline
Worklaod           & \multicolumn{6}{c|}{TPC-H}                        & \multicolumn{6}{c}{TPC-DS}        \\ \hline
Method             & mean & 50th & 90th & 95th & 99th & max    & mean & 50th & 90th  & 95th  & 99th & max  \\ \hline
GCN                & 1.63 & 1.61 & 1.95 & 2.04 & 2.22 & 2.75   & 3.48 & 1.97 & 7.8   & 8.67  & 9.49 & 10.02 \\
TCN                & 1.60 & 1.57 & 1.93 & 1.99 & 2.19 & 2.64   & 3.48 & 1.67 & 7.08  & 8.60  & 12.95& 415.7 \\
QPPNet             & 2.75 & \textbf{1.48} & 3.24 & 5.27 & 24.73& 517.85 & 3.91 & 3.28 & 7.69  & 7.82  & 8.09 & 8.57  \\
QueryFormer        & 1.60 & 1.51 & 2.08 & 2.17 & 2.37 & 69.58  & 2.59 & 2.02 & 4.88  & 6.75  & 8.61 & 12.14   \\
RAAL               & 2.24 & 2.31 & 2.63 & 2.82 & 3.85 & 5.13   & 2.00 & \textbf{1.43} & 3.18  & 3.86  & 5.23 & 5.54 \\ \hline
CONCERTO             & \textbf{1.46} & 1.49 & \textbf{1.90} & \textbf{1.98} & \textbf{2.05} & \textbf{2.49}   & \textbf{1.84} & 1.73 & \textbf{2.78}  & \textbf{2.88}  & \textbf{3.22} & \textbf{3.84}   \\ 
CONCERTO w/o OCP     & 1.70 & 1.61 & 2.46 & 3.13 & 3.26 & 3.35   & 2.05 & 1.57 & 3.62  & 5.26  & 8.27 & 9.49  \\
CONCERTO w/o ResAttn & 2.09 & 1.97 & 3.05 & 3.25 & 4.29 & 6.52   & 2.48 & 1.90 & 4.95  & 5.62  & 6.68 & 7.72 \\ \hline
\end{tabular}
\end{table*}

\begin{figure*}[htbp]
  \centering
  \includegraphics[width=0.9\linewidth]{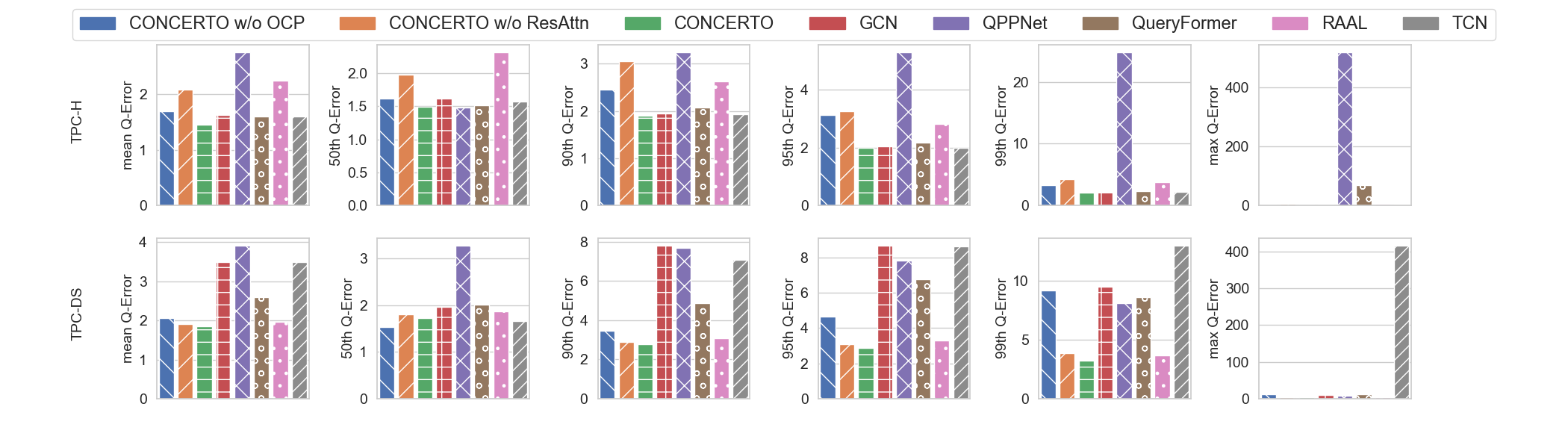}
  \vspace{-15pt}
  \caption{Q-Error distribution comparison.}
  \vspace{-15pt}
  \label{fig:accuracy_comp}
\end{figure*}

The results of CONCERTO's and baselines' accuracy are shown in \autoref{tab:accuracy} and \autoref{fig:accuracy_comp}. On both TPC-H and TPC-DS workloads, CONCERTO significantly outperforms all baselines. The Q-Error distribution shows consistent performance gains across OLAP workloads of varying size and complexity. The TCN's Q-Error is quite low on TPC-H but increases a lot on TPC-DS, especially the max Q-Error. We also noticed that both GCN and TCN can reach similar accuracy compared to CONCERTO on TPC-H, but their Q-Error gains a lot on TPC-DS. This proves the effectiveness of these two widely used methods. But it also indicates that when it comes to complex OLAP workloads on high-performance DBMSs, such methods lose their accuracy due to a lack of targeted design. On the other hand, CONCERTO maintains a stable lead on both workloads. The QPPNet's long-tail distribution problem on TPC-H is indeed serious. We have tried everything that we could, such as adjusting its hyperparameters and switching to better loss functions like LeakyReLU and Swish. This is the best result we can get. We also noticed that, although TPC-DS is a much more complex workload, the max Q-Error of QPPNet was significantly reduced compared to TPC-H. 

We think the problem is that QPPNet is not a resource-aware approach that cannot model parallel pipeline execution well in certain situations. With enough training templates, it may perform better due to the ability to generalize. However, TPC-H is a workload that only has a few templates, and 23.5\% templates used for testing is too high for them. Therefore, when there are serious resource competitions in some queries, QPPNet, which is trained on limited TPC-H workloads, cannot predict the latency accurately. This result provides indirect evidence that traditional learning methods are difficult to adapt to high-performance OLAP DBMSs, and the resource-aware mechanism is crucial in them.

\begin{figure}[htbp]
  \centering
  \includegraphics[width=\linewidth]{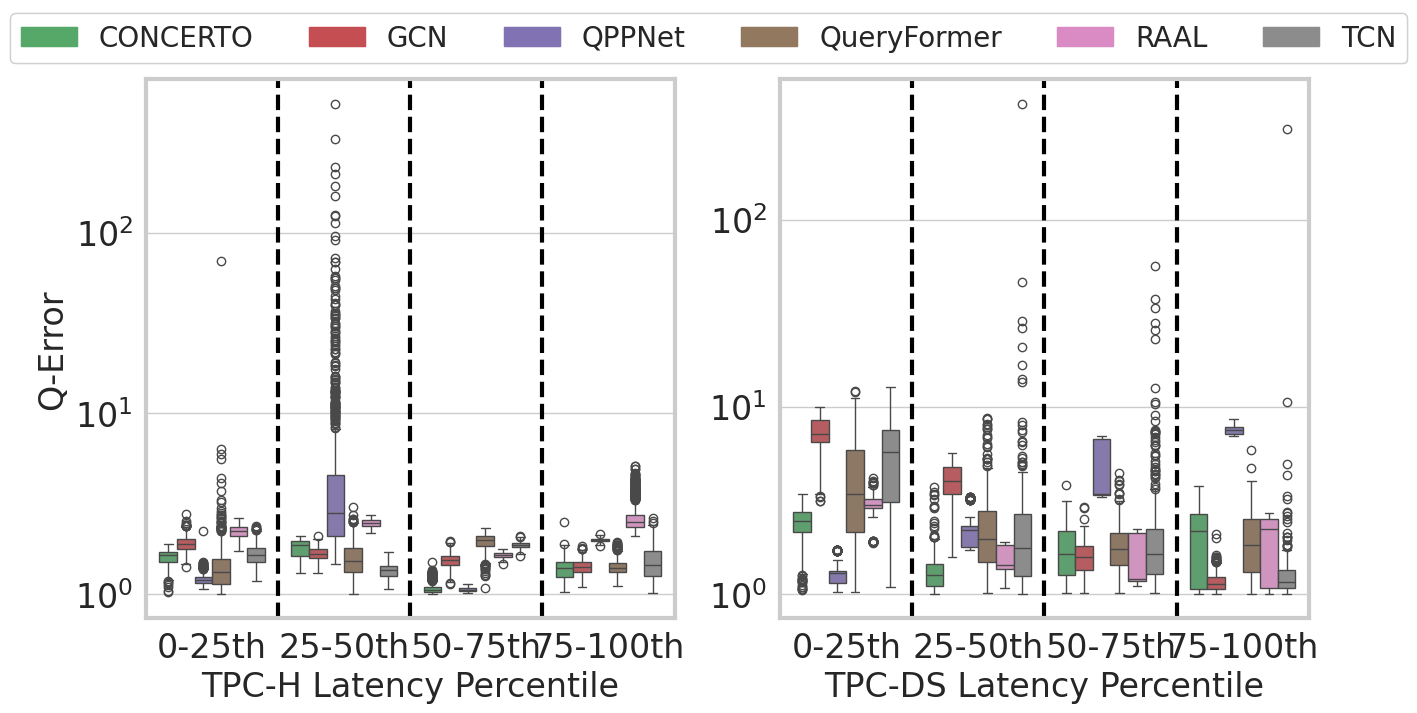}
  \caption{Q-Error distribution grouped by queries' latency. Box plots are drawn in the order of the legend.}
  \vspace{-5pt}
  \label{fig:Quant_Dist}
\end{figure}

In order to more intuitively compare the error distribution between different methods, we grouped the methods by latency range and plotted the accuracy box plots for each method, since the latency can reflect the complexity of the query. The results are shown in \autoref{fig:Quant_Dist}. The results show that the Q-Error of the QPP task in different query execution time groups does not show an obvious increasing trend. We think this is because the accuracy of query-driven QPP methods is more affected by the generalization between different templates than by the size of the query plan. The long tail distribution of methods such as QPPNet and TCN is concentrated in a certain group interval because the execution time of queries generated by the same query template is more concentrated than that of queries generated by different templates. This phenomenon occurs when the QPP method has poor generalization performance on individual templates. Taking the QPPNet as an example, its Q-Error has a significant peak at the 25- 50th group. As analyzed above, the number of training templates is too few and the testing templates' ratio is too high, so QPPNet cannot handle resource competition between operators well on individual templates. So, most of the queries generated by the specific template fall into the same interval and cause this result. For resource-aware methods like CONCERTO and RAAL, a relatively stable Q-Error can be maintained under different query complexities, that is, their long-tail distribution problem is relatively mild. This reveals that in high-performance OLAP DBMS, the usage of various hardware resources in the system and the resource competition of operators are crucial to the accuracy of QPP tasks. 

\textbf{Ablation Study.}
By comparing the accuracy of ablation methods CONCERTO w/o OCP and CONCERTO w/o ResAttn, we can conclude that the resource attention mechanism has the biggest improvement in the prediction accuracy, and both modules can significantly improve the accuracy.

\subsection{Performance Comparison}

\begin{figure}[htbp]
  \captionsetup{skip=5pt} 
  \centering
  \includegraphics[width=0.7\linewidth]{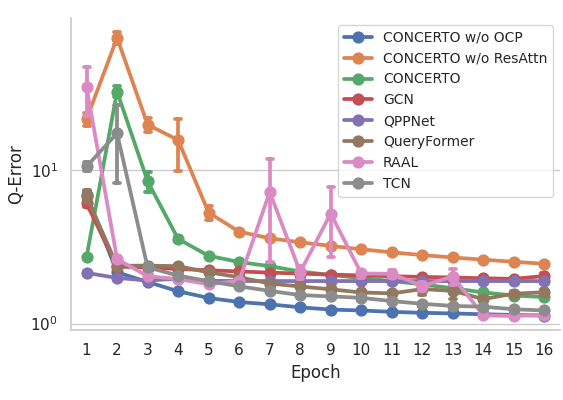}
  \caption{Q-Error with a standard error during training.}
  \label{fig:tpch_ConvergenceCurve}
\end{figure}

The training speed, inference latency, and model size are crucial to the practicality of the QPP approach. We report the performance of CONCERTO and compare it with baselines as shown in \autoref{tab:performance}. Since TCN is a component of CONCERTO, we didn't include it for comparison. For the performance of training speed, the QueryFormer is the fastest because it uses a well-optimized Transformer architecture. Since CONCERTO employs multi-stage models and dynamic data-flow tree expression, which involves a re-indexing of tensors. So CONCERTO is a little bit slower than QueryFormer and becomes the second fastest method among all compared ones. QPPNet suffers from its dynamic network structure, significantly slowing down its training speed. RAAL uses the most complex network structure, and training takes 599.5 seconds per epoch. The training speed itself can not reflect the actual training efficiency, so we draw their convergence curve in \autoref{fig:tpch_ConvergenceCurve}. The QPPNet converges the fastest, which takes 5 epochs to drop from 2.14 to 1.89. But its average Q-Error is quite high compared with other methods. Its accuracy on the test set shows similar results. However, as the methods that have the lowest Q-Error on the train set, RAAL and CONCERTO w/o OCP have higher Q-Error than CONCERTO on the test set, which shows their poorer generalization ability. In general, most methods reach their lowest Q-Error at around the 14th epoch. The difference in convergence is much significant than their training time cost per epoch.

In terms of model size and inference speed, we summarized each approach's model size by counting each parameter's size obtained by the \textit{state\_dict()}. Note that all OCP models' parameters are counted in. As shown in \autoref{tab:performance}, CONCERTO has the smallest model and fastest inference latency among all methods. RAAL suffers from its recurrent neural network mechanism, which can not perform parallel inference and has the highest inference latency. QPPNet needs to build very large networks to match the complex OLAP queries, thus, it has the highest inference latency.

\begin{table}[htbp]
\centering
\caption{Performance comparison of baselines, CONCERTO. The data is from models trained on the TPC-H workload.}
\label{tab:performance}
\begin{tabular}{l|lll}
\hline
Method             & Size (MB) & Training (s/epoch) & Inference (ms) \\  \hline
QPPNet             & 0.103     & 531.4              & 303       \\
QueryFormer        & 0.414     & \textbf{53.4}      & 7.1        \\
RAAL               & 0.121     & 599.5              & 64.0         \\ 
CONCERTO             & \textbf{0.094} & 78.4          & \textbf{4.2}          \\  
\hline
\end{tabular}
\end{table}

\vspace{-8pt}
\subsection{Robustness to Cardinality Errors}
\begin{figure}[htbp]
  \centering
  \includegraphics[width=\linewidth]{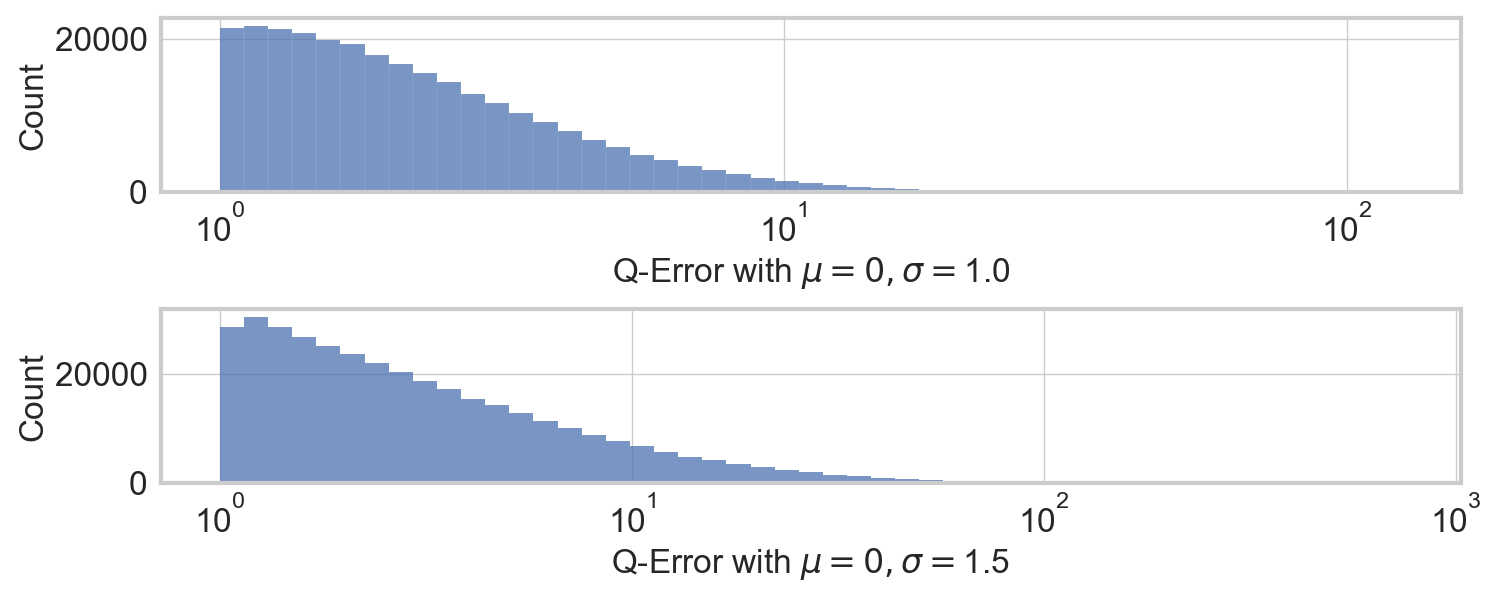}
  \caption{Distribution of generated Q-Error with log-scale x-axis. The Q-Error's maximum value is 740.08.}
  \vspace{-5pt}
  \label{fig:noise_dist}
\end{figure}

The cardinality errors usually have a significant impact on cost estimation and query performance prediction tasks, as the input scale directly decides the cost. Therefore, we evaluate CONCERTO's robustness to cardinality errors in this subsection. Due to the complexity of ClickHouse's pipeline builder and dynamic executor, it is hard to adapt multiple cardinality estimation methods to its kernel. Instead, we generate Q-Errors that follow a log-normal distribution and simulate the estimation result by multiplying it with the true cardinality collected by CONCERTO's \textsf{Runtime Tracker}. As shown in \autoref{fig:noise_dist}, the Q-Error distribution of generated cardinality is similar to real estimation with a long-tailed distribution.

We assess CONCERTO's accuracy as the change of the Q-Error distribution on TPC-H workloads. Due to QPPNet's accuracy being significantly lower than other baselines, we did not include it in the experiment. We set log-normal distribution's $\mu=0$ and the standard deviation $\sigma=\{1.0, 1.5\}$.

\begin{table}[htbp]
\centering
\caption{Robustness of Cardinality Error, measured by mean Q-Error.}
\label{tab:robustness}
\begin{tabular}{l|lll}
\hline
Method             &  True Card   & $\sigma=1.0$  & $\sigma=1.5$ \\  \hline
QueryFormer        &  1.60        &  1.59         & \textbf{1.61}   \\
RAAL               &  2.24        &  2.27         & 2.30    \\ 
CONCERTO             & \textbf{1.46}&\textbf{1.49}  & 2.59  \\  
\hline
\end{tabular}
\vspace{-5pt}
\end{table}

\begin{figure}[!htbp]
  \centering
  \captionsetup{skip=5pt} 
  \includegraphics[width=\linewidth]{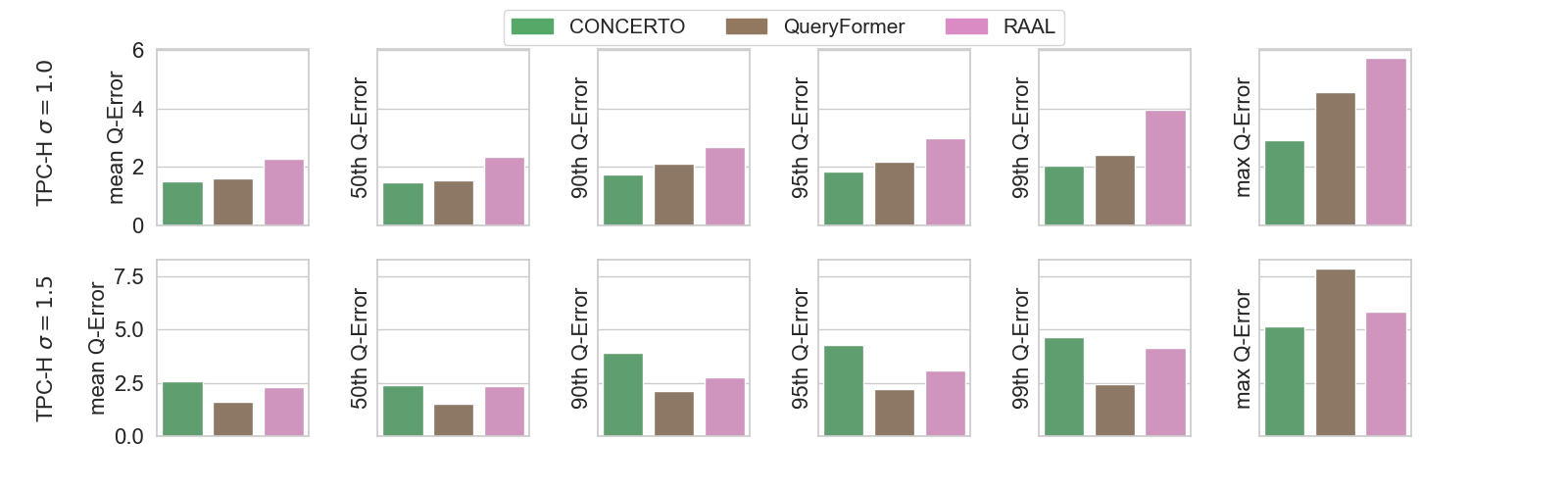}
  \caption{Robustness Q-Error distribution comparison.}
  \label{fig:card_error_comp}
\end{figure}

As \autoref{tab:robustness} and \autoref{fig:card_error_comp} show, the predicted query performance's mean and max Q-Error increase as the $\sigma$ of Q-Error distribution increases, as we expected. In terms of max Q-Error, CONCERTO outperforms baselines on each setting of $\sigma$. Although QueryFormer's mean Q-Error almost doesn't increase, its max Q-Error increases much faster than CONCERTO and RAAL. But in terms of mean Q-Error, CONCERTO's is higher than the two baselines when $\sigma=1.5$. This shows that although CONCERTO continues to benefit from the insignificant long-tail distribution caused by its design, its robustness remains to be improved under larger cardinality estimation errors.

\vspace{-6pt}
\section{Conclusion and Future Works}
In this paper, we propose a novel query performance prediction approach, CONCERTO, for high-performance OLAP DBMSs with complex query execution mechanisms like vectorized operators, DAG execution plan, and dynamic parallel pipeline. By carefully tracking and collecting data with the \textsf{Runtime Tracker} and employing a multi-stage resource-competition-aware model, CONCERTO can capture the resource competition and dynamic pipeline modifications better in such DBMSs. We implemented CONCERTO for ClickHouse and adapted three baselines for comparison, The experimental results show that CONCERTO outperforms all baselines in accuracy, model size, and inference speed, while its training speed and robustness are also practical. We plan to expand CONCERTO and create a cross-DBMS query performance prediction plugin that can automatically adapt to different DBMSs.

\section*{Acknowledgments}
This work is supported by the National Natural Science Foundation of China (NSFC) (62232005).


 





\end{document}